\documentclass[preprint,12pt]{elsarticle}

\usepackage{amssymb}
\usepackage{amsmath}
\usepackage{algorithm}
\usepackage{algpseudocode}
\usepackage{booktabs}
\usepackage{float}
\usepackage{graphicx}
\usepackage{hyperref}
\usepackage{tikz}
\usetikzlibrary{arrows.meta, positioning, shapes.geometric}

\journal{Journal of Computational Physics}

\newcommand{\ntheta}{n_\theta}
\newcommand{\nphi}{n_\phi}
\newcommand{\Anorm}{A_{\rm norm}}

\begin{document}

\begin{frontmatter}

\title{Mosaic: Area-Closed Spherical Surface Mosaics Induced by Cartesian Grids}

\author[labelMath]{H.F. Counts}
\author[labelMath]{Mihaela Dobrescu}
\author[labelSEC,labelJLAB]{D. Heddle}
\author[labelSEC]{Aubrie Kooiker}
\author[labelMath]{Walter Pierce}

\affiliation[labelMath]{organization={Department of Mathematics, Christopher Newport University},
            addressline={1 Avenue of the Arts},
            city={Newport News},
            postcode={23606},
            state={Virginia},
            country={USA}}

\affiliation[labelSEC]{organization={School of Engineering and Computing, Christopher Newport University},
            addressline={1 Avenue of the Arts},
            city={Newport News},
            postcode={23606},
            state={Virginia},
            country={USA}}

\affiliation[labelJLAB]{organization={The Thomas Jefferson National Accelerator Facility},
            addressline={12000 Jefferson Avenue},
            city={Newport News},
            postcode={23606},
            state={Virginia},
            country={USA}}

\begin{abstract}
We describe \textit{Mosaic}, a computational geometry method for constructing the surface mosaic induced when a Cartesian volume grid intersects a spherical shell. The motivating application is conservative coupling between data produced on rectangular grids and diagnostics or boundary conditions defined on spherical surfaces, as occurs in space-weather, magnetohydrodynamic, atmospheric, and geophysical models. The method identifies Cartesian cells that intersect the shell, constructs cell--sphere prepatches, splices those regions by the spherical colatitude grid, and then splices by azimuth to produce final patches indexed by the five-tuple $(n_x,n_y,n_z,\ntheta,\nphi)$. The implementation explicitly treats the polar singularity by separating polar-derived theta patches from ordinary phi splicing. A near-pole numerical failure mode, caused by linear interpolation in azimuth, is removed by computing exact intersections between great-circle boundary segments and meridian planes. The prepatch construction also handles several degeneracy cases that occur beyond the ordinary corner-straddling geometry, including doubly-crossing edges, lens-shaped prepatches, secondary closed loops, and re-entrant face arcs. For a representative nonuniform Cartesian test grid, the current implementation builds 3618 intersecting Cartesian cells, 3602 ordinary prepatches, 6476 theta patches, and 9714 final phi patches, with all 3602 ordinary cells built successfully and zero theta or phi splicing failures. The final phi-spliced patches close in normalized area to roundoff relative to their theta-spliced parents. The method is implemented in the open-source Java/Maven application \texttt{mdi-mosaic}, which provides visualization, diagnostics, and JSON export of final patch boundaries and normalized areas.
\end{abstract}

\begin{keyword}
Grid intersection \sep Spherical geometry \sep Space weather modeling \sep
Magnetohydrodynamics \sep Computational geometry \sep Scientific visualization
\MSC 65M50 \sep 65N50 \sep 86A10
\end{keyword}

\end{frontmatter}

\section{Introduction}

Numerical simulations of physical systems often use different coordinate systems for different parts of a calculation. Cartesian grids are convenient for structured finite-volume and finite-difference methods in Euclidean domains \cite{LeVeque2002}, while spherical grids are natural for planetary, atmospheric, astrophysical, and geophysical boundaries \cite{Lauritzen2010}. Coupling data between these descriptions can be difficult when conservation matters. A simple interpolation from one grid to another may be adequate for visualization, but it does not by itself provide the geometric areas needed for conservative flux transfer.

This problem appears in space-weather modeling, where global plasma transport may be computed on structured or adaptively refined grids while quantities of interest are evaluated on spherical shells centered on a planet \cite{White1998,Toth2012,gombosi2021space}. Related issues arise in magnetohydrodynamics \cite{Dedner2002}, atmospheric modeling on spherical or cubed-sphere grids \cite{Putman2007}, astrophysical simulation \cite{Stone2008}, and geophysical fluid dynamics \cite{Ringler2013}. Overset or Chimera grids \cite{Benek1986} provide one common strategy for communicating between overlapping coordinate patches, while adaptive mesh refinement \cite{Berger1984} provides local resolution where needed. The Mosaic approach described here is complementary: it computes the induced surface partition directly.

The contribution here is narrower and more geometric than a complete coupling framework such as the Space Weather Modeling Framework. Mosaic does not provide a global simulation infrastructure, adaptive mesh management, or interpolation of physical fields. Instead, it constructs the surface partition induced by the intersection of a Cartesian volume grid, a spherical shell, and a spherical angular grid. The output is a set of explicit boundary curves and normalized areas for the induced surface patches. This makes the method suitable as a geometric preprocessing or diagnostic step for conservative coupling: field interpolation, flux averaging, and conservation policies can be built on top of the exported patch geometry.

Given a Cartesian grid enclosing a sphere of radius $R$ and a spherical angular grid on that sphere, Mosaic constructs a set of surface patches. Each final patch belongs to one Cartesian cell and one spherical angular cell, and is labeled by
\begin{equation}
  (n_x,n_y,n_z,\ntheta,\nphi).
\end{equation}
The exported quantities include the boundary vertices, a normalized area
\begin{equation}
  \Anorm = \frac{A}{4\pi R^2},
\end{equation}
and a dimensionless perimeter $L/R$. For a complete normalized partition, the mathematical target is
\begin{equation}
  \sum_i \Anorm^{(i)} = 1.
\end{equation}
The numerical diagnostics below report deviations from this value, or deviations from the parent patch area when testing an intermediate splicing stage.

\section{Grid definitions}
The implementation used for the results below assumes a common Cartesian coordinate system, with the sphere centered at the origin. This is the natural setting for the current GSM-coordinate application, where GSM denotes Geocentric Solar Magnetospheric coordinates: the origin is at Earth center, the $x$ axis points approximately toward the Sun, the $z$ axis is chosen from the projection of Earth's magnetic dipole axis, and the $y$ axis completes the right-handed system. The underlying geometric operations, however, are local to the cell and sphere and do not require uniform grid spacing.

A Cartesian cell is denoted by $(n_x,n_y,n_z)$. A spherical angular cell is denoted by $(\ntheta,\nphi)$. At the poles, azimuth is singular; the implementation therefore permits polar aggregate patches with $\nphi$ recorded as an aggregate value rather than as an ordinary azimuth bin.

\section{Algorithm overview}
\label{AlgorithmOverview}

The current implementation uses four main stages.

\begin{enumerate}
\item \textbf{Find intersecting Cartesian cells.} Cells whose interiors or faces intersect the spherical shell are identified and classified.
\item \textbf{Build ordinary prepatches.} For ordinary corner-straddling cells, the intersection of the Cartesian cell with the sphere is represented as a closed loop of cell-face/sphere boundary curves. Each prepatch is labeled by $(n_x,n_y,n_z)$.
\item \textbf{Theta splice.} Each prepatch is clipped by the spherical colatitude grid, producing theta patches labeled by $(n_x,n_y,n_z,\ntheta)$.
\item \textbf{Phi splice.} Each theta patch is clipped by the azimuth grid to produce final patches labeled by $(n_x,n_y,n_z,\ntheta,\nphi)$. Polar-derived theta patches are handled by a separate polar phi-splicing path.
\end{enumerate}

\begin{figure}[H]
\centering
\resizebox{0.6\textwidth}{!}{%
\begin{tikzpicture}[
    font=\small,
    node distance=0.95cm,
    >=Latex,
 process/.style={
    rectangle,
    rounded corners,
    draw=black,
    thick,
    align=center,
    text width=8.5cm,
    minimum height=0.95cm,
    inner sep=5pt,
    execute at begin node={\hyphenpenalty=10000\exhyphenpenalty=10000\relax}
},
    arrow/.style={
        ->,
        thick
    }
]

\node[process] (input) {
    Cartesian volume grid and spherical shell
};

\node[process, below=of input] (step1) {
    Step 1: Identify shell-intersecting Cartesian cells\\
    \emph{typical run: 3618 intersecting cells}
};

\node[process, below=of step1] (step2) {
    Step 2: Build ordinary prepatch boundary curves\\
    \emph{3602 ordinary prepatches; face-penetration cases classified}
};

\node[process, below=of step2] (step3) {
    Step 3: Splice prepatches by $\theta$\\
    \emph{6476 theta patches; polar prepatches handled}
};

\node[process, below=of step3] (step4) {
    Step 4: Splice theta patches by $\phi$\\
    \emph{ordinary phi splicer plus polar phi splicer}
};

\node[process, below=of step4] (output) {
    Final spherical mosaic patches\\
    \emph{9714 final patches; area closure to roundoff}
};

\node[process, below=of output] (export) {
    Visualization, diagnostics, and JSON export
};

\draw[arrow] (input) -- (step1);
\draw[arrow] (step1) -- (step2);
\draw[arrow] (step2) -- (step3);
\draw[arrow] (step3) -- (step4);
\draw[arrow] (step4) -- (output);
\draw[arrow] (output) -- (export);

\end{tikzpicture}%
}
\caption{Flow diagram of the MDI Mosaic construction algorithm. The method begins with a Cartesian grid and a spherical shell, identifies shell-intersecting cells, constructs prepatch boundaries, splices those patches first in $\theta$ and then in $\phi$, and finally produces an area-closed set of spherical mosaic patches suitable for visualization and export. Representative diagnostic counts from the test configuration are included in each stage.}
\label{fig:algorithm-flow}
\end{figure}
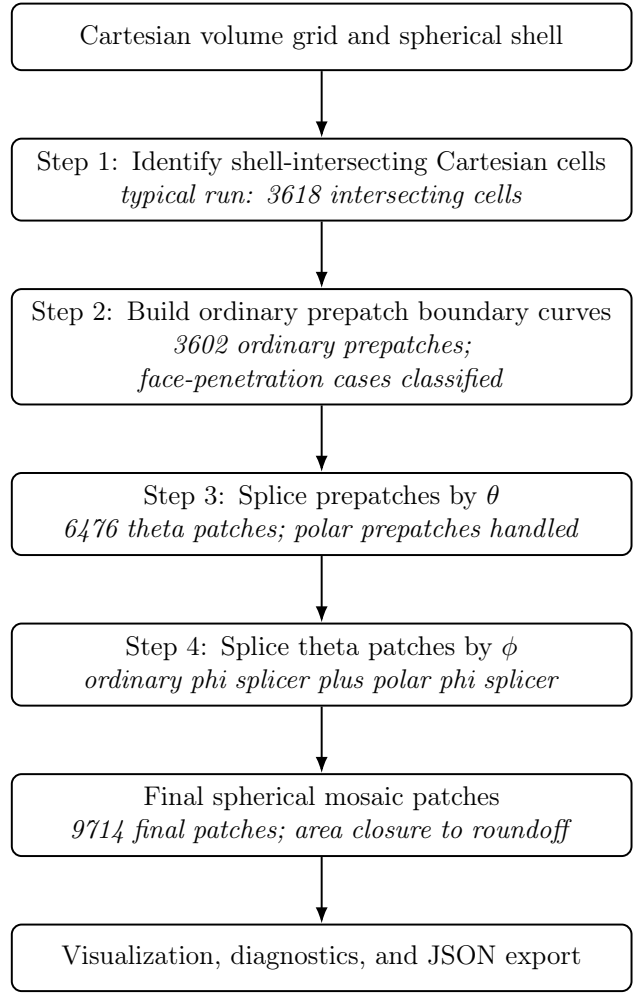

The result is a set of final patches whose areas can be accumulated, exported, and used for conservative surface coupling. Figure~\ref{fig:algorithm-flow} summarizes the current processing pipeline.

\section{Intersecting cells}
\label{IntersectionCells}

The first stage identifies Cartesian cells that intersect the sphere. For most cells this is a corner-straddling test: at least one corner lies inside the sphere and at least one lies outside. Such cells contain one or more cell edges that intersect the sphere and normally produce ordinary prepatches.

A special case occurs when a cell face penetrates the sphere even though all cell corners are outside. These face-penetration cells have no inside corners and may have no ordinary edge-crossing signature. The implementation detects them by comparing the closest point on a cell face to the sphere center. If this closest point lies within the face and has distance less than $R$, the cell is classified as a face-penetration cell. If the projection falls outside the face, the closest point on the face boundary is used instead.

For the representative test grid in Sec.~\ref{ResultsSec}, the algorithm identifies 3618 intersecting cells. Of these, 3602 are ordinary corner-straddling cells and 16 are classified as face-penetration cells with no inside corners. The production pipeline builds ordinary prepatches for the 3602 corner-straddling cells and attempts to build prepatches for the 16 face-penetration cells via a dedicated face-penetration builder (Section~\ref{PrepatchSec}). For this grid, 8 of the 16 face-penetration cells produce valid prepatches; the remaining 8 are deferred as genuinely degenerate cases that do not affect area closure at the diagnostic precision used here.

\section{Prepatch construction}
\label{PrepatchSec}

A prepatch is the portion of the sphere cut out by a single Cartesian cell, before the spherical angular grid is applied. It is labeled by the Cartesian cell index $(n_x,n_y,n_z)$.

For an ordinary intersecting cell, each intersected cell face contributes a curve lying simultaneously on the spherical surface and on the plane of that face. Geometrically, this is an arc of the circle formed by intersecting the sphere with the face plane. The endpoints of such an arc occur where cell edges intersect the sphere. The face arcs are then connected in order to form a closed boundary loop.

\begin{figure}[H]%% placement specifier
\centering%% For centre alignment of image.
\includegraphics[scale=0.4]{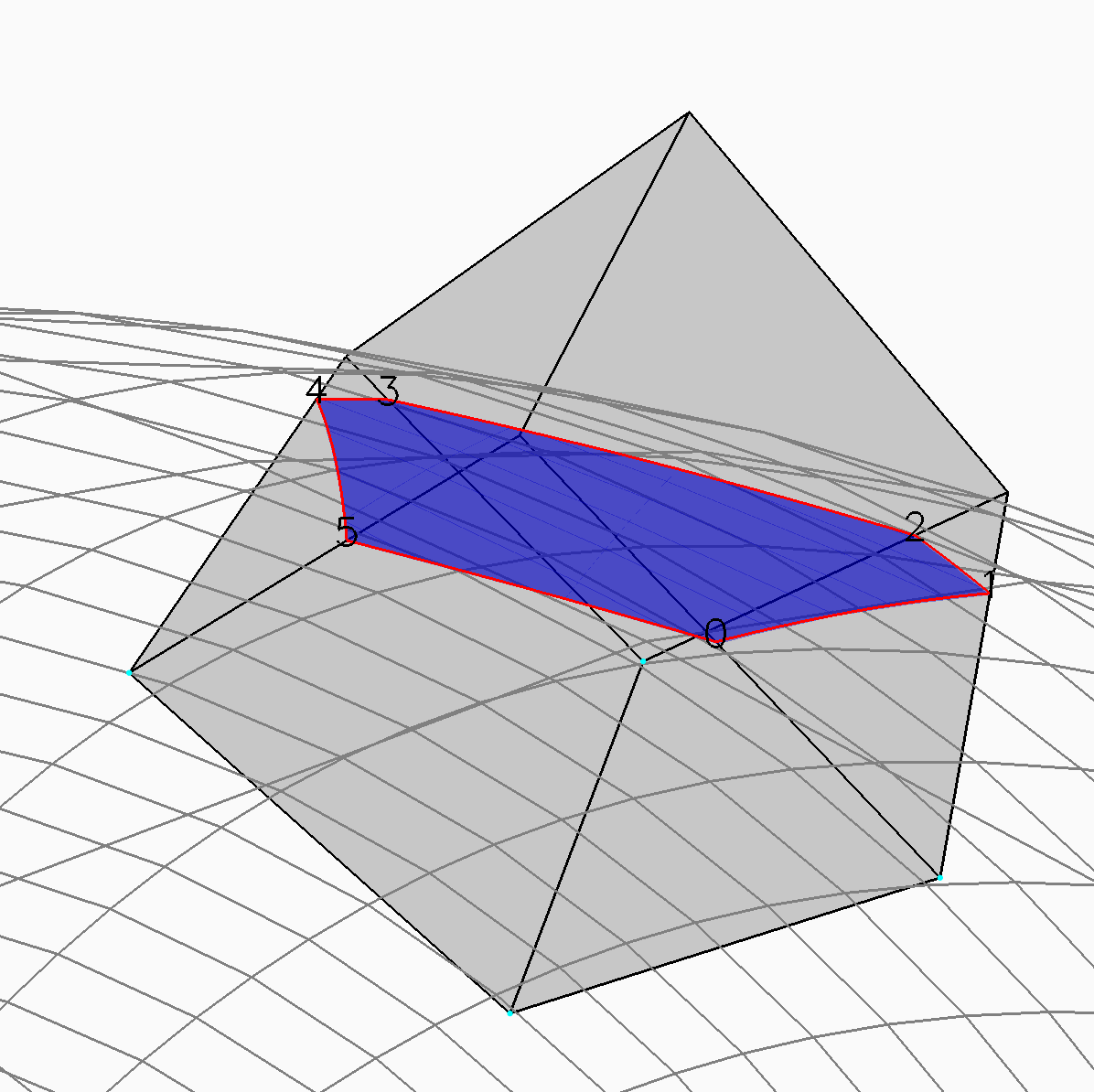}
\caption{An ordinary prepatch from an intersection with four corners inside the sphere and six intersecting edges.}
\label{Prepatch}
\end{figure}

A previous implementation \cite{Pierce2013Chimera} described the face arcs through an explicit rotated-coordinate construction. The current implementation is organized around the same geometry but uses a more direct boundary representation suitable for subsequent clipping and visualization. The important invariant is that each ordinary prepatch boundary (Figure \ref{Prepatch}) is a closed loop on the sphere, with a consistent orientation and with enough sampled or canonicalized points to support area diagnostics and later theta/phi splicing.

\subsection{Geometric degeneracy cases}
\label{DegeneracySec}

Testing on multiple grids revealed four geometric cases beyond the ordinary corner-straddling case that must be handled to achieve complete area closure.

\paragraph{Doubly-crossing edges} The original edge-sphere intersection test used a sign-straddle condition: an edge is crossed if one endpoint is inside the sphere and one is outside ($f_0 \cdot f_1 < 0$, where $f = \|\mathbf{p}\|^2 - R^2$). This misses edges where both endpoints are outside the sphere ($f_0 > 0$, $f_1 > 0$) but the edge interior dips inside, producing two intersection points. Such edges arise in cells near the equatorial plane of the grid where the sphere surface cuts horizontally across a cell face. Detecting them requires solving the full quadratic $\|\mathbf{p}(t)\|^2 = R^2$ along the edge and checking that both roots lie strictly in the interior $(0,1)$.

\paragraph{Lens-shaped prepatches} When a cell has no straddle edges but contains a doubly-crossing edge, the sphere surface intersects the cell in a closed lens shape bounded by two arcs: one on each of the two cell faces adjacent to the doubly-crossing edge. The prepatch builder detects this case (exactly two intersection points from the same edge) and builds the two-arc closed boundary directly.

\paragraph{Secondary loops in straddle cells} A corner-straddling cell may also contain doubly-crossing edges that produce a second, disconnected closed loop on the sphere within the same cell. Adding these intersection points to the main prepatch's curve list corrupts the curve-chaining topology. The implementation instead builds the secondary loop as a separate prepatch sharing the same Cartesian cell identifier, after the main straddle prepatch is built successfully.

\paragraph{Re-entrant face arcs} For cells near the boundary of the sphere's intersection with the grid, the face circle on a given face may dip outside the face rectangle between the two intersection points. In this case neither candidate arc lies fully within the face rectangle, causing the scored arc-selection to fail. The implementation falls back to \texttt{createShorterArc}, which bypasses face-rectangle scoring and selects the shorter of the two arcs unconditionally. For the tested configurations, this fallback selects the geometrically correct boundary segment: the shorter arc that traces the sphere surface within the cell.

\subsection{Face-penetration cells}

Face-penetration cells have all eight corners outside the sphere but have a face through which the sphere penetrates. The sphere intersects the face plane in a circle of radius $r_c = \sqrt{R^2 - d^2}$, where $d$ is the distance from the origin to the face plane. The \texttt{FacePenetrationPrepatchBuilder} handles three sub-cases: the face circle lies entirely within the face rectangle (full-circle prepatch), the circle is clipped by the face rectangle on one side (two-crossing clip), and the circle is clipped on two sides (four-crossing clip).

\section{Theta splicing}
\label{ThetaSpliceSubsection}

Theta splicing clips each prepatch by the colatitude grid. Let the spherical grid lines be $\theta_m$. The goal is to subdivide each prepatch into regions that lie within a single theta band,
\begin{equation}
  \theta_m \leq \theta \leq \theta_{m+1}.
\end{equation}
Each resulting theta patch is labeled by $(n_x,n_y,n_z,\ntheta)$.

The theta splicer inserts intersections between prepatch boundaries and theta-grid curves and then reconstructs closed child boundaries within each theta band. Polar prepatches require special care because the spherical coordinate system degenerates at $\theta=0$ and $\theta=\pi$. In the current implementation, prepatches involving the poles are classified and handled by a fan-like decomposition so that theta patches are produced without leaving polar holes.

For the test grid, theta splicing visits 3602 prepatches and builds 6476 theta patches. No prepatches fail in this stage. The theta-spliced area differs from the ordinary prepatch diagnostic reference by approximately $1.36\times 10^{-7}$ in normalized area. This small excess is localized in near-pole, high-curvature cells and is retained as the reference for the final phi-splicing area closure tests.

\section{Phi splicing and polar handling}
\label{PhiSpliceSubsection}

Phi splicing clips each theta patch by azimuthal grid lines. Away from the poles, this produces ordinary final patches labeled by $(n_x,n_y,n_z,\ntheta,\nphi)$.

A key architectural decision in the current implementation is to separate polar-derived theta patches from ordinary phi splicing. These patches are passed to a dedicated polar phi splicer, which produces polar aggregate patches. The aggregate representation avoids forcing a coordinate-singular region through ordinary azimuthal clipping. In the exported JSON, such patches are marked as polar aggregates and have no ordinary $\nphi$ value.

\subsection{Near-pole meridian intersections}

The most important numerical correction in the current implementation occurs in ordinary phi splicing near the poles. A previous version found the intersection of a boundary segment with a meridian by interpolating linearly in azimuth. This is unreliable near a pole, where $\phi$ changes rapidly and is ill-conditioned. The error was small in absolute area but visible in aggregate diagnostics: two near-pole nonpolar theta patches accounted for most of the final area loss.

The corrected method computes the intersection geometrically. Let $\mathbf{a}$ and $\mathbf{b}$ be unit vectors at the endpoints of a boundary segment, represented for the purpose of the final patch boundary as a short great-circle segment. The great-circle plane has normal
\begin{equation}
  \mathbf{g} = \mathbf{a} \times \mathbf{b}.
\end{equation}
The meridian at azimuth $\phi_c$ lies in the plane
\begin{equation}
  -\sin\phi_c\,x + \cos\phi_c\,y = 0,
\end{equation}
with normal
\begin{equation}
  \mathbf{m}=(-\sin\phi_c,\cos\phi_c,0).
\end{equation}
The candidate intersection direction is therefore parallel to
\begin{equation}
  \mathbf{q}=\mathbf{g}\times\mathbf{m}.
\end{equation}
The sign of $\mathbf{q}$ is chosen so that the point lies on the requested meridian ray, and the candidate is accepted only if it lies on the short arc from $\mathbf{a}$ to $\mathbf{b}$. This replaces a coordinate interpolation with a coordinate-free plane intersection. With this change, the ordinary nonpolar phi area error for the test grid falls to floating-point roundoff.

\section{Area and perimeter diagnostics}
\label{AreaSec}

The implementation uses normalized area,
\begin{equation}
  \Anorm = \frac{A}{4\pi R^2},
\end{equation}
as the primary area quantity. This makes the output independent of the physical radius scale and provides a simple global closure diagnostic. The final exported JSON records normalized area for each patch and a total normalized area for the mosaic.

Patch areas are computed from canonicalized boundary point sequences on the sphere. The boundary curves produced by the geometric construction are represented by sufficiently resolved point sequences for splicing, rendering, and diagnostics. The area of each child region is evaluated by connecting adjacent boundary points with spherical (great circle) segments and accumulating the corresponding spherical polygon area. This convention is also used by the final JSON export, which records boundary vertices and dimensionless perimeter $L/R$.

The perimeter diagnostic is likewise exported in dimensionless form. If $\alpha_i$ is the central angle between consecutive boundary vertices, then
\begin{equation}
  \frac{L}{R} = \sum_i \alpha_i.
\end{equation}
For applications requiring physical area or perimeter, the conversions are
\begin{equation}
  A = 4\pi R^2\Anorm, \qquad L = R\sum_i \alpha_i.
\end{equation}

\section{Representative test grid and results}
\label{ResultsSec}

For the current regression test, we use an axis-aligned nonuniform Cartesian grid enclosing a sphere of radius
\begin{equation}
  R = 4.60993.
\end{equation}
The spherical grid uses 48 uniform colatitude cells and 32 uniform azimuth cells.
\begin{table}[H]
\centering
\begin{tabular}{ll}
\toprule
Quantity & Value \\
\midrule
Coordinate system & GSM (Geocentric Solar Magnetospheric) \\
Origin & Earth center \\
Length unit & Earth radii, $R_E$ \\
Sphere radius & $R = 4.60993\,R_E$ \\
Cartesian $x$ range & $[-5.5,\,5.5]\,R_E$ \\
Cartesian $y$ range & $[-5.66670,\,5.68030]\,R_E$ \\
Cartesian $z$ range & $[-5.5,\,5.5]\,R_E$ \\
Cartesian grid cells & $33 \times 34 \times 33$ \\
Theta cells & 48 \\
Phi cells & 32 \\
Theta spacing & Uniform in $\theta$ \\
Phi spacing & Uniform in $\phi$ \\
\bottomrule
\end{tabular}
\caption{Definition of the representative test grid. Coordinates are expressed in GSM coordinates with distances measured in Earth radii.}
\label{tab:test-grid}
\end{table}

Table~\ref{tab:counts} summarizes a representative run.

\begin{table}[H]
\centering
\begin{tabular}{lr}
\toprule
Quantity & Count \\
\midrule
Intersecting Cartesian cells & 3618 \\
Ordinary corner-straddling cells & 3602 \\
Face-penetration cells detected & 16 \\
Face-penetration prepatches built & 8 \\
Deferred face-penetration cells & 8 \\
Ordinary prepatches built & 3602 \\
Theta patches built & 6476 \\
Final phi patches built & 9714 \\
Polar-derived theta patches handled & 234 \\
Polar final aggregate patches & 6 \\
\bottomrule
\end{tabular}
\caption{Representative patch counts for the current test grid.}
\label{tab:counts}
\end{table}

Table~\ref{tab:area} summarizes the area diagnostics. The phi-spliced final patches close against their theta-parent reference to roundoff. The remaining difference between the theta-spliced reference and the original ordinary prepatch diagnostic reference is a separate theta-stage diagnostic, not a phi-splicing loss.

\begin{table}[H]
\centering
\small
\begin{tabular}{@{}p{0.40\linewidth}p{0.25\linewidth}p{0.25\linewidth}@{}}
\toprule
Diagnostic & Normalized area & Difference \\
\midrule
Ordinary prepatch diagnostic reference & 0.9999999999999989 & --- \\
Theta patches & 1.0000001362237638 & $+1.362\times 10^{-7}$ \\
Final phi patches & 1.0000001362237634 & $\approx -7.7\times 10^{-15}$ vs. theta reference \\
Nonpolar phi patches & 0.9991669903659882 & $\approx -2.1\times 10^{-15}$ vs. nonpolar theta reference \\
Polar phi patches & 0.0008331458577753 & $\approx +2.2\times 10^{-19}$ vs. polar theta reference \\
\bottomrule
\end{tabular}
\caption{Representative normalized-area diagnostics. The phi stage closes to floating-point roundoff relative to the theta-spliced parent reference.}
\label{tab:area}
\end{table}

\subsection{Grid-sensitivity and Degeneracy Study}
\label{GridSensitivitySec}

To test whether the area-closure behavior is specific to the representative grid, we run the same pipeline on several related grid configurations. These tests are intended as grid-sensitivity checks rather than a formal convergence study for a differential equation. The key quantity is the final phi-stage area difference measured relative to the theta-spliced parent reference; robust phi splicing should preserve this reference to roundoff across changes in Cartesian and angular resolution. The results for five grids of varying coarseness are given in Table~\ref{tab:grid-sensitivity}.

\begin{table}[H]
\centering
\small
\begin{tabular}{@{}llrrrrr@{}}
\toprule
Case & Cartesian cells & $\theta$ cells & $\phi$ cells
& Final patches & $1-\Anorm$ & $\Delta A_\phi$ vs. theta ref. \\
\midrule
A & $33\times34\times33$ & 48 & 32 & 9714   & $1.1\times10^{-15}$  & $-7.7\times10^{-15}$ \\
B & $24\times35\times24$ & 48 & 32 & 8640   & $5.6\times10^{-16}$  & $1.3\times10^{-15}$ \\
C & $40\times41\times40$ & 48 & 32 & 13,407 & $3.6\times10^{-15}$  & $3.1\times10^{-15}$ \\
D & $33\times34\times33$ & 36 & 24 & 8450   & $-4.2\times10^{-5}$  & $1.8\times10^{-15}$ \\
E & $33\times34\times33$ & 64 & 48 & 14,088 & $-4.1\times10^{-5}$  & $5.4\times10^{-6}$ \\
\bottomrule
\end{tabular}
\caption{Grid-sensitivity tests. The column $1-\Anorm$ reports the deviation of
the final total normalized area from unity. The final column reports the
normalized-area difference between the final phi-spliced patches and their
theta-spliced parent reference. Thus $1-\Anorm$ measures global closure of the
exported mosaic, while $\Delta A_\phi$ isolates the area change introduced by
the final phi-splicing stage.}
\label{tab:grid-sensitivity}
\end{table}
We do not claim a formal convergence rate from these tests. Rather, the
purpose is to separate global mosaic closure from the area preservation of the
final phi-splicing stage. Cases A--C close globally to roundoff and also preserve
the theta-spliced parent reference to roundoff. Cases D and E, which change the
spherical angular resolution, show small but visible global normalized-area
deviations of order $10^{-5}$. These deviations are associated primarily with
earlier-stage boundary reconstruction and theta-splicing effects, not with the
ordinary Cartesian-cell intersection test. In case D the final phi stage still
preserves its theta-parent reference to roundoff. In case E, the higher angular
resolution exposes a residual phi-stage difference of order $10^{-6}$, indicating
that angular-resolution sensitivity remains a useful target for further
diagnostics.

The implementation includes diagnostic overlays that display prepatches, theta patches, final patches, polar final patches, and mouse-over patch identifiers. These diagnostics were essential for detecting the near-pole phi interpolation error and for confirming that apparent polar holes in an orthographic projection were visualization artifacts rather than missing area.

\section{Visualization and Monte Carlo diagnostics}
\label{VisualizationSec}

Monte Carlo sampling is used in this work as an independent visualization and diagnostic tool. Random points on the sphere are assigned to both a Cartesian cell and a spherical angular cell, producing a colored map of the induced mosaic. This sampling does not replace the deterministic construction: it does not provide exact boundary curves, reliable detection of small or nearly tangential regions, or accurate patch areas. It is nevertheless useful as a visual cross-check of the deterministic patch boundaries, especially when overlaid with the final patch outlines.

\begin{figure}[H]
\centering
\includegraphics[scale=0.45]{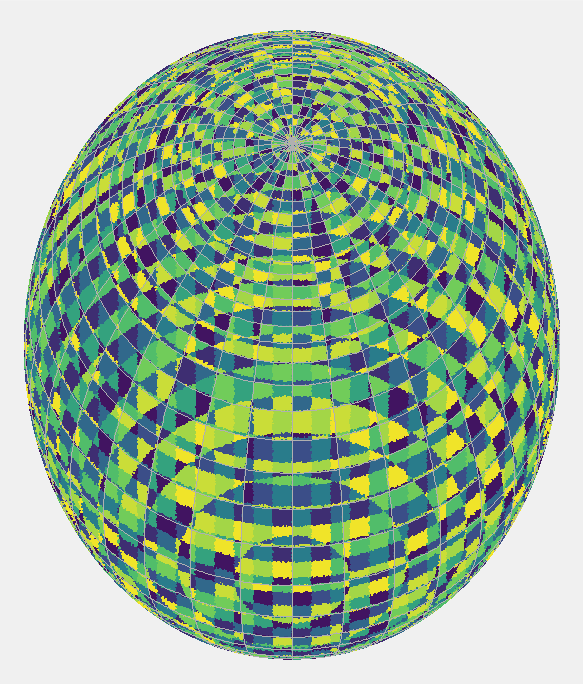}
\caption{Monte Carlo visualization of the representative test grid using 20 million random points on the sphere. The colors indicate the cell assignments obtained by sampling rather than by deterministic boundary construction.}
\label{fig:monte-carlo}
\end{figure}

\section{JSON export}
\label{JsonExportSec}

The final patch list can be exported as JSON. The export is intended for downstream analysis rather than as a compact visualization format. Each patch record includes its identifier, parent Cartesian cell, theta index, phi index or aggregate marker, polar-aggregate flag, normalized area, dimensionless perimeter, and boundary vertices. The schema records the convention
\begin{equation}
  \texttt{normalizedArea} = A/(4\pi R^2),
\end{equation}
and
\begin{equation}
  \texttt{perimeterOverRadius}=L/R.
\end{equation}

A typical export for the test grid contains 9714 final patch records. The compact export is substantially smaller than a fully verbose export containing both Cartesian and spherical boundary representations and per-segment records, but it can still be tens of megabytes.

\section{Implementation}
\label{ImplementationSec}

The current implementation is the Java/Maven application \texttt{mdi-mosaic}. 
The source code is available at \url{https://github.com/heddle/mdi-mosaic}, and the results reported here were produced with release \texttt{v1.0}: 
\url{https://github.com/heddle/mdi-mosaic/releases/tag/v1.0}. 
It is built on the MDI scientific visualization framework~\cite{HeddleMDI2026} and uses an MDI map view for projection, overlays, mouse feedback, and algorithm diagnostics.
The application includes the algorithm pipeline, graphical controls for running and clearing results, diagnostic overlays, final-patch mouse-over lookup, and JSON export.

\begin{figure}[H]%% placement specifier
\centering%% For centre alignment of image.
\includegraphics[scale=0.4]{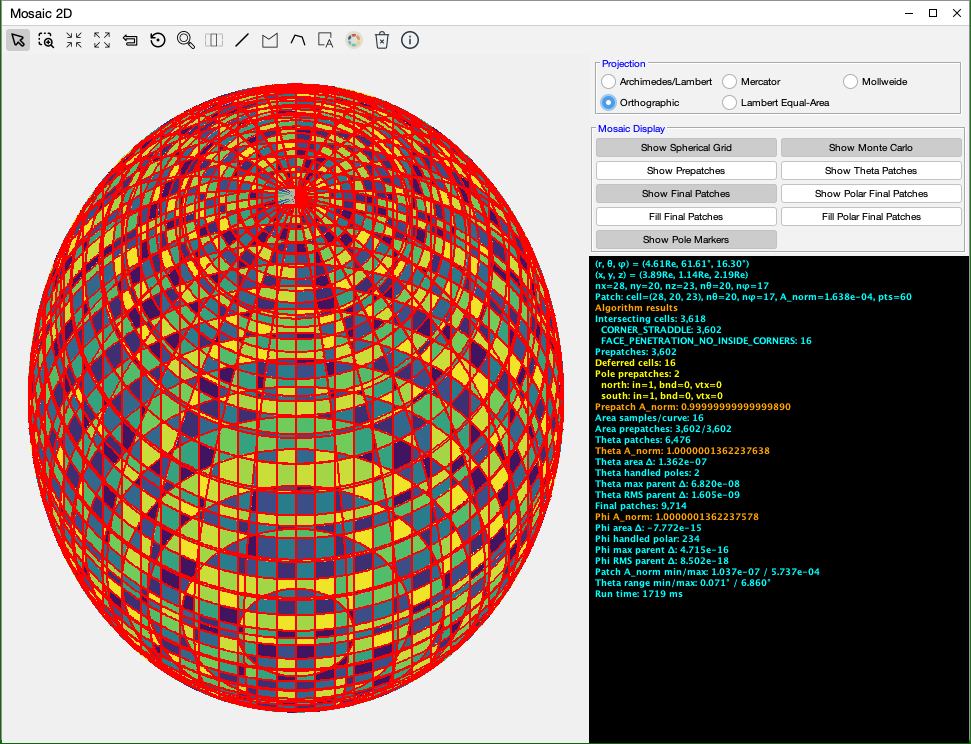}
\caption{The MDI Mosaic application with final patches and mouse-over patch diagnostics. The patch coloring comes from the Monte Carlo diagnostic overlay. As expected, the final patch boundaries, shown in red, form perimeters around the Monte Carlo colorization.}
\label{fig:mdi-mosaic-app}
\end{figure}

\section{Limitations and future work}

The current implementation handles the ordinary corner-straddling cases, doubly-crossing edge cases, lens-shaped prepatches, secondary loops, re-entrant face arcs, and several face-penetration sub-cases. A small number of genuinely degenerate face-penetration configurations remain deferred.
Several items remain for future work.

First, 8 of the 16 face-penetration cells on the representative test grid are still deferred as genuinely degenerate cases. For this grid they do not affect area closure at the diagnostic precision used, but a fully general production implementation should construct and validate patches for all face-penetration sub-cases explicitly.

Second, the theta-splicing stage introduces a small normalized-area excess of order $10^{-7}$ for the representative grid. The evidence from parent-level diagnostics suggests that this is associated with boundary reconstruction and spherical-polygon area evaluation in near-pole, high-curvature cells. The phi stage preserves the theta reference to roundoff, but a future implementation should investigate whether the theta-stage excess can be reduced by exact conic intersections or improved boundary reconstruction near the poles.

Third, the current JSON export is intentionally geometric. Coupling this mosaic to a physical simulation requires additional interpolation or averaging of field quantities from the parent Cartesian cells and a conservation policy for fluxes or extensive quantities.

Fourth, the grid-sensitivity study in Table~\ref{tab:grid-sensitivity} should be expanded to include coarser and finer Cartesian grids, different angular resolutions, and sphere placements that exercise a wider range of face-penetration and near-tangent configurations.

\section{Conclusion}

Mosaic constructs the spherical surface partition induced by a Cartesian grid and a spherical angular grid. The current algorithm proceeds through intersecting-cell detection, ordinary prepatch construction, theta splicing, and phi splicing with separate polar handling. Two robustness improvements are described. The first is the use of exact great-circle/meridian intersections during ordinary phi splicing, which removes a near-pole area-loss mode caused by linear interpolation in azimuth. The second is the treatment of four geometric degeneracy cases in prepatch construction:
doubly-crossing edges, lens-shaped prepatches, secondary loops in straddle cells,
and re-entrant face arcs. These cases were previously unhandled and caused area loss on grids other than the representative test case. With these improvements, all 3602 ordinary corner-straddling cells on the representative test grid produce valid prepatches with zero theta or phi splicing failures, and the final phi patches close in normalized area to roundoff relative to the theta-spliced parent reference. Grid-sensitivity tests on five configurations show roundoff-level global and phi-stage closure for the Cartesian-resolution cases, while angular-resolution
changes expose smaller residual boundary-reconstruction and phi-splicing
effects that remain targets for further refinement.

\section*{Acknowledgments}
This work was supported in part by NSF grant 2415038.

\bibliographystyle{elsarticle-num}
\bibliography{references}

\end{document}